\documentclass[a4paper,12pt]{article}
\usepackage{normal}

\begin{document}

\newcommand{\ket}[1]{\ensuremath{|#1 \rangle}}
\newcommand{\bra}[1]{\ensuremath{\langle #1|}}
\newcommand{\braket}[2]{\ensuremath{\langle #1|#2 \rangle}}
\newcommand{\ketbra}[2]{\ensuremath{|#1 \rangle \langle #2|}}
\newcommand{\ro}[1]{\ensuremath{|#1 \rangle \langle #1|}}
\newcommand{\av}[1]{\ensuremath{\langle #1 \rangle}}

\newcommand{\real}{\ensuremath{\mathrm{Re}}}
\newcommand{\trace}{\ensuremath{\textsf{Tr}}}

\newcommand{\id}{\ensuremath{\mathsf{1}}}
\newcommand{\iden}{\ensuremath{{\sf 1\hspace*{-1.0ex}\rule{0.15ex}
{1.2ex}\hspace*{1.0ex}}}}
\newcommand{\R}{\ensuremath{{\sf R\hspace*{-0.9ex}\rule{0.15ex}
{1.5ex}\hspace*{0.9ex}}}}
\newcommand{\N}{\ensuremath{{\sf N\hspace*{-1.0ex}\rule{0.15ex}
{1.3ex}\hspace*{1.0ex}}}}
\newcommand{\Q}{\ensuremath{{\sf Q\hspace*{-1.1ex}\rule{0.15ex}
{1.5ex}\hspace*{1.1ex}}}}
\newcommand{\C}{\ensuremath{{\sf C\hspace*{-0.9ex}\rule{0.15ex}
{1.3ex}\hspace*{0.9ex}}}}

\newcommand{\h}[1]{\ensuremath{\mathcal{H}_{#1}}}

\newcommand{\me}{\ensuremath{\mathrm{e}}}
\newcommand{\mi}{\ensuremath{\mathrm{i}}}

\newcommand{\de}{\ensuremath{\mathrm{d}}}
\newcommand{\dd}[2]{\ensuremath{\frac{\mathrm{d}#1}{\mathrm{d}#2}}}
\newcommand{\ddd}[2]{\ensuremath{\frac{\mathrm{d}^2#1}{\mathrm{d}#2^2}}}

\newcommand{\ot}[2]{\ensuremath{\left( \begin{array}{c} #1 \\ #2
\end{array} \right)}}
\newcommand{\oth}[3]{\ensuremath{\left( \begin{array}{c} #1 \\ #2 \\ #3
\end{array} \right)}}
\newcommand{\twtw}[4]{\ensuremath{\left( \begin{array}{cc} #1 & #2 \\
#3 & #4 \end{array} \right)}}
\newcommand{\thth}[9]{\ensuremath{\left( \begin{array}{ccc} #1 & #2 & #3
\\ #4 & #5 & #6 \\ #7 & #8 & #9 \end{array} \right)}}

\newcommand{\expp}[1]{\ensuremath{\me^{\mi\hat{H}#1}}}
\newcommand{\expm}[1]{\ensuremath{\me^{-\mi\hat{H}#1}}}

\title{The Many Worlds of Uncertainty}
\author{Clare Hewitt-Horsman\\ \small{Theoretical Quantum Optics Group,}\\\small{Imperial College, London}}
\maketitle

\section{Introduction}

The status of the uncertainty relations varies between the different interpretations of quantum mechanics. At one
end of the scale, for Heisenberg they showed the fundamental limits of applicability of the wave and particle
pictures separately. At the other end, in hidden variable theories they are merely statistical mechanical relations
of no foundational significance. The aim of the current paper is to explore their meanings within a certain
neo-Everettian many worlds interpretation. We will also look at questions that have been linked with the
uncertainty relations since Heisenberg's uncertainty principle: those of joint and repeated measurement of
non-commuting (or otherwise `incompatible') observables. This will have implications beyond the uncertainty
relations, as we will see the fundamentally different way in which statistical statements are interpreted in the
neo-Everett theory that we use.

\section{Neo-Everettian Many Worlds}

It is firstly necessary briefly to describe the interpretation with which we will explore the uncertainty
relations. This has been described (with full references) in \cite{me},
and is a variant of the many worlds theory started by Everett \cite{everett}.\\

The salient points are these. As with all Everett-style theories, the measurement problem is solved by having all
possible outcomes of an interaction instantiated. The `worlds' of which there are `many' are not defined by a
precise rule, but rather are structures identified ``for all practical purposes''. They are structures marked out
by their usefulness, explanatory relevance and persistence over time, as high-order rather than fundamental
ontology. They occur at all levels: there are structures of decoherent branches that persist over large timescales
and act in ways that we
would call `a universe', and within them can be identified other substructures.\\

 To take the old example, within
the state of a Schro\"{o}dinger Cat experiment we would be able to identify two structures that persisted stably
over time: one of a cat alive (and unbroken apparatus) and one of a dead cat (and associated apparatus). These are
not, of course, the only structures identifiable in that state (we need only to change basis), but they persist in
a stable state over timescales relevant to the experiment, and indeed after the diabolical device is opened they
become elements in the decoherence basis. The identification of the cats is not precise: just as, in everyday life,
we chose to call an imprecisely defined and constantly changing bundle of atoms `a cat', so when we look at the
quantum state of the system we see something that looks like that which we would normally refer to as `an alive
cat', and something that looks like a dead cat. We therefore say we have two worlds, one in which there is an alive
cat, and one in which there is a dead one\footnote{These worlds at this point only extend as far as the bounds of
the apparatus - the theory is local.}.\\

One unusual feature of this type of many worlds theory is that, while at larger scales decoherence picks out the
relevant structures for us, at very small scales there can be more than one basis in which we find relevant
structures. This will, of course, depend on what we are looking for and how the system is evolving -- again, the
identification of structures is not fundamental.

\section{The Uncertainty Relations}\label{sec}

Before coming on to the uncertainty relations that we will be looking at, it is worth saying with what we will
\emph{not} be dealing.\\

We will not be using the `textbook' form of the uncertainty relations, the Robertson relation for non-commuting
observables:
\begin{equation} \Delta A \ \Delta B \ge \frac{1}{2} |\av{[A,B]}| \label{rob}\end{equation}

\noindent where
$$ \Delta A = \av{A^2} - \av{A}^2 \ \ \ \ \ \  \ \ \Delta B = \av{B^2} - \av{B}^2$$

There are various standard problems with this formulation (summarized in \cite{onlineurs})\footnote{The stronger
relation owing to Schr\"{o}dinger,
$$ (\Delta A)^2(\Delta B)^2 \ge \frac{1}{4}|\av{[A,B]}|^2 + \frac{1}{4} \av{\{ (A - \av{A}), (B - \av{B})\} }^2$$
\noindent will also not be dealt with, as it suffers from the same problems.}. For present purposes the main
problems are:
\begin{itemize}
\item These relations tell us nothing when one of the observables is in an eigenstate as then the right hand side
of the inequality is zero. There are also other situations when they give no information -- for example, in a
spin-half system, because $[S_x,S_z] = \mi \hbar S_y$, if $S_y=0$ then the relations tell us nothing about $S_x$
and $S_z$. This is even though they are non-commuting observables of the system, which are considered the objects
of the uncertainty relations on this view. They are not, therefore, fully general inequalities.
\item The relations work only for Gaussian (or near-Gaussian) distributions. The interpretation of the quantities
$\Delta A$ and $\Delta B$ as spreads only works for such distributions.
\end{itemize}

Neither will we look at any uncertainty relations including time -- given the status of time these will be very
different objects from other uncertainty relations. Lastly, we will not be looking at entropic uncertainty
relations, for the very reason that Deutsch \cite{dd} champions them: they deal not with fundamentals (which we
wish to
look at to analyse from a neo-Everettian perspective) but with operational situations.\\

We will be using the uncertainty relations owing to Uffink \cite{uffinktwonew}:
\begin{equation} \delta \theta \ W(\hat{A}) \ge
\hbar \ \mathrm{arcos} \left(\frac{1 + \beta - \alpha}{\alpha}\right)\label{ur}\end{equation}

The quantities are defined as follows: $\hat{A}$ is an operator that generates a group of states
$\ket{\psi_\theta}$ by
$$\ket{\psi_\theta} = \me^{\frac{\mi \theta \hat{A}}{\hbar}} \ \ket{\psi}$$
\noindent $\delta \theta$ is the smallest value of $\theta^\prime$ such that
$$ |\braket{\psi_{\theta}}{\psi_{\theta + \theta^\prime}}| = \beta < 1$$
\noindent $W(\hat{A})$ is the smallest interval in the distribution $\braket{\psi}{a}$ (where $\hat{A}\ket{a} =
a\ket{a}$) in which a fraction $\alpha < 1$ is contained:
$$ \int_W |\braket{\psi}{a}|^2 \mathrm{da} = \alpha$$
\noindent Finally it is a requirement of the relation that $\beta \ge 2\alpha - 1$.\\

If we chose $\beta$ such that states become more-or-less distinguishable once their overlap reaches it, the
quantity $\delta \theta$ is therefore the expected error in estimated a given state: states in the range
$\ket{\psi_{\theta - \delta \theta}}$ to $\ket{\psi_{\theta + \delta \theta}}$ are indistinguishable by any
measurements. \\

The `minimal interpretation' \cite{onlineurs} of these identities is that they are relations between constraints on
different probability distributions of the same state, for groups and their generators\footnote{The most famous of
these in the context of uncertainty relations are position and momentum, each of which acts as the generator of the
other group. The set of spacial translations is generated by
$$ 1 - \frac{\mi}{\hbar} \bf{p}.d\bf{x^\prime}$$
\noindent and that of translations in momentum space by
$$1 - \frac{\mi}{\hbar} \bf{x}.d\bf{p^\prime}$$}. They do not fully define the distributions, but give conditions they must necessarily follow. $\delta \theta$
gives a constraint on the $\braket{\psi}{\psi_\theta}$ probability distribution: the more spread out the
probability distribution, the less sharply we would be able to estimate the
state, even given multiple measurements on an ensemble of similarly prepared states.\\

To go beyond this minimal interpretation in the direction of our many worlds theory we will need to make an
assumption about probability. The interpretation of probability is an enormous problem for Everett-style theories,
and one which we will not make any attempt to solve here. We will instead make the (hopefully minimal) assumption
that probabilities are attached in a basic way to worlds. For whichever part of a state we take to be a single
`world', the square of the associated coefficient is the probability given to that world. Beyond this, we keep
silence\footnote{It is worth pointing out that in our neo-Everett view there is no ontological classical
probability: all such probability arises because of quantum probabilities. We therefore cannot say, for example,
``We are entangled with a given world but we do not know which one'': if we cannot tell {\bf by any means} which of
a
bunch of worlds we are entangled with then its relative state is a superposition of those worlds.}.\\

The standard interpretation of a probability distribution in quantum mechanics is as the distribution of the
outcomes of measurements on an ensemble of similarly prepared states. Given this, the relations (\ref{ur}) are
statistical relations for an ensemble of states, and therefore give only a statistical relation for an individual
system.  However, in our neo-Everettian interpretation we do not need to invoke an ensemble to make sense of our
statistics as we have one within a single state -- an ensemble of worlds and their associated probabilities.
Probability distributions are therefore distributions of probability over \emph{worlds}, and refer to single systems.\\

For an example, take a single qubit system in the state $\frac{1}{\sqrt{2}}(\ket{\uparrow} + \ket{\downarrow})$.
Standardly the probability distribution would be described as something like: ``do measurements in the $\{
\ket{\uparrow},\ket{\downarrow}\}$ basis on a large number of similarly prepared systems,  and roughly 50\% will be
up and 50\% down.''. In the neo-Everett theory it refers ontologically to the fact that within the single qubit
system itself can be identified two worlds in the $\{ \ket{\uparrow},\ket{\downarrow}\}$ basis, both of which have
associated probability 0.5. The former description is, of course, still how we would gain epistemic access to the
distribution: we can of course only access one world at a time by measurement.\\

A probability distribution is therefore a set of worlds and their associated probabilities. The uncertainty
relations (\ref{ur}) refer to two such sets that can be identified within the same state -- the
$\ket{\psi_\theta}$-worlds and the $\ket{a}$-worlds, both of which sets are found in the (single) state
$\ket{\psi}$. The relations constrain what, given any set of $\ket{a}$-worlds, the spread of the associated set of
$\ket{\psi_\theta}$-worlds is, and \emph{vice-versa}. They are not in one sense fundamental: for any state we could
look at the states of a group generator $\hat{A}$ and the states of the group and work out $\delta \theta$,
$W(\hat{A})$ and we would find that they satisfy (\ref{ur}). However the relations do tell us more than that: that
we would find this for \emph{any} group and its generator in \emph{any} state.\\

Fundamentally, the relations (\ref{ur}) for a neo-Everettian say that it is impossible physically to have a state
whose worlds corresponding to groups and their generators have probability distributions the spreads of which
violate (\ref{ur}). Furthermore, this is the kind of interpretation that is given to \emph{all} statistical
statements in neo-Everett theory: they are concerning the worlds of a single system and their associated
probabilities, which is very different from the standard interpretation. As we shall see below, this will lead to
physical conclusions different from the standard interpretation, as probabilistic statements constrain individual
systems and not just ensembles.

\section{Measurements}

\subsection{Joint Measurement}

\begin{quotation} \emph{The uncertainty principle refers to the degree of indeterminateness in the possible present
knowledge of the simultaneous values of various quantities with which the quantum theory deals...} \end{quotation}

\noindent Ever since these fateful words of Heisenberg \cite[p20]{hei} (and the subsequent notorious microscope
exposition), the uncertainty relations have been regarded in large portions of the literature to be fundamentally
statements about the joint measurement of non-commuting observables. ``You can't measure position and momentum
simultaneously" is a very common statement and, more generally, ``you can't simultaneously measure two
non-commuting observables exactly". Such conclusion are almost invariably drawn from inequalities such as
(\ref{rob}) and pose a number of problems\footnote{quite apart from -- as was pointed out in \S\ref{sec} -- the
silence of (\ref{rob}) on
matters where one or other of the states is an eigenstate.}.\\

To begin with, such a statement is, in point of fact, false. It is fairly uncontroversial that if there is a
simultaneous eigenstate of two quantities then on any reasonable theory of measurement we will be able to perform a
measurement giving us these quantities. Non-commuting observables do not share \emph{complete} sets of simultaneous
eigenkets, but nothing stops them sharing \emph{some}. For example \cite[p33]{sakuri}, a system with angular
momentum $l=0$ is in a simultaneous
eigenstate of $L_x=0$ and $L_z=0$, even though they do not commute.\\

Furthermore, and fundamentally, the relations (\ref{rob}) (and indeed our relations (\ref{ur})) are independent of
a theory of measurement -- they are relations pertaining to statistical distributions. Until a connection is made
between the statistics of many measurements and the state of a system in a singular, given measurement situation,
these relations can tell us nothing about what is or is not allowed in one measurement. Indeed, on the `standard'
view of the relations the distributions referred to are gained by two \emph{separate} sets of multiple measurements
on similarly prepared states, there is not a joint measurement in sight.\\

Our neo-Everett theory does, however, give us the connection between statistics and singular states -- \emph{and} a
measurement theory\footnote{A measurement of the first kind (state preparation), in the $\{0,1\}$ basis, would be
written:
$$ \ket{\mathrm{Ready}} \otimes (a\ket{0} + b\ket{1}) \rightarrow a\ket{R(0)}\otimes \ket{0} +
b\ket{R(1)}\otimes\ket{1}$$ \noindent In words: at the start of the experiment we identify a structure that is a
measuring device set to `ready', and a quantum state in a superposition of 0 and 1. Afterwards we see two
structures (`worlds'), one where the readout of the device is `0' and the state of the system under measurement is
\ket{0}, and one where they are `1' and \ket{1}.} -- and thus we are allowed to try and draw some inferences from
the uncertainty relations within it.\\

We concluded in the previous section that the uncertainty relations tell us what set of $\ket{a}$-worlds
corresponds to a given set of $\ket{\psi_\theta}$-worlds (and \emph{vice-versa}), for the worlds of a single
system. We can therefore see that the set of, say, $\ket{a}$-worlds that corresponds to a single state
$\ket{\psi_\theta}$ cannot be an eigenstate of $\ket{a}$; that is, within the state in which we can identify a
single $\ket{\psi_\theta}$-world we identify many $\ket{a}$-worlds. This is true for \emph{all} single
$\ket{\psi_\theta}$- and $\ket{a}$- worlds: there is never a structure in which we can identify only one
$\ket{\psi_\theta}$-world and one $\ket{a}$-world.\\

The implications of this for the question of joint measurement are fairly straightforward. A measuring device
cannot couple to a world which is both a given $\ket{\psi_\theta}$-world and an eigenvalue of $\hat{A}$ because
such a world does not exist. Such a coupling would be what is meant in our neo-Everett interpretation by
`measurement', so we conclude that a joint measurement of $\theta$ and $\hat{A}$ is not possible.\\

We are able in this case, unlike with the relations (\ref{rob}), to be sure that our conclusion is universally
valid; as \cite{uffinktwonew} points out, (\ref{rob}) depends on the actual state being used, but (\ref{ur}) does
not. This means we do not have the problems that (\ref{rob}) has -- there is no possibility of joint
eigenstates or of the relation being meaningless in some states.\\

The question now is how much of this comes from the uncertainty relations themselves. The simple fact that the
states are not jointly measurable comes from their lack of shared eigenkets, and there will be many other states
and observables that are not jointly measurable but which are not linked by uncertainty relations. What we can say
is that the uncertainty relations relate a sub-set of states which are not jointly measurable to the underlying
group structure that connects them. Without the relations they would be disparate phenomena, and the content of the
relations is in relating them. Lack of joint measurement is a consequence (in a neo-Everettian framework) of the
group structure, but not all states which are not jointly measurable are groups and their generators.

\subsection{Repeated Measurements}

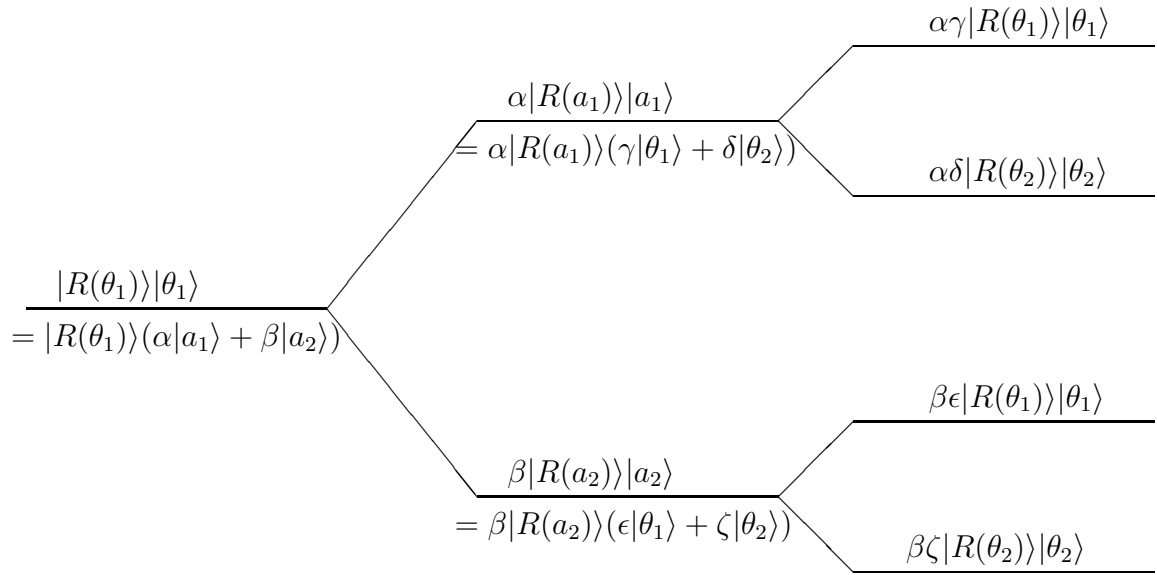
\begin{figure} \caption{\label{figure} The many worlds of a repeated measurement}

\setlength{\unitlength}{1cm}
\begin{picture}(15,11)

\put(0,6){\line(1,0){4}} \put(0.4,6.2){$\ket{R(\theta_1)} \ket{\theta_1}$}
\put(-0.2,5.5){$=\ket{R(\theta_1)}(\alpha\ket{a_1}+\beta\ket{a_2})$}

\put(4,6){\line(4,5){2}} \put(4,6){\line(4,-5){2}}

\put(6,8.5){\line(1,0){4}} \put(6.4,8.7){$\alpha\ket{R(a_1)}\ket{a_1}$}
\put(5.7,8){$=\alpha\ket{R(a_1)}(\gamma\ket{\theta_1} + \delta\ket{\theta_2})$}

\put(6,3.5){\line(1,0){4}} \put(6.4,3.7){$\beta\ket{R(a_2)}\ket{a_2}$}
\put(5.7,3){$=\beta\ket{R(a_2)}(\epsilon\ket{\theta_1} + \zeta\ket{\theta_2})$}

\put(10,8.5){\line(1,1){1}} \put(10,8.5){\line(1,-1){1}} \put(10,3.5){\line(1,1){1}} \put(10,3.5){\line(1,-1){1}}

\put(11,4.5){\line(1,0){4}} \put(12,4.7){$\beta\epsilon\ket{R(\theta_1)}\ket{\theta_1}$}

\put(11,2.5){\line(1,0){4}} \put(11.7,2.7){$\beta\zeta\ket{R(\theta_2)}\ket{\theta_2}$}

\put(11,9.5){\line(1,0){4}} \put(12,9.7){$\alpha\gamma\ket{R(\theta_1)}\ket{\theta_1}$}

\put(11,7.5){\line(1,0){4}} \put(12,7.7){$\alpha\delta\ket{R(\theta_2)}\ket{\theta_2}$}

\end{picture}

\end{figure}

We now move on to a situation that has been linked to the uncertainty relations (although to a lesser extent that
joint measurements): repeated measurements. In this situation we perform a $\theta$-measurement, then an
\emph{a}-measurement, and then a $\theta$-measurement. As is well known, we will find that the final measurement of
$\theta$ bears no relation to the initial. This is often taken to be a consequence of the uncertainty relations:
that the two distributions cannot be equally sharp and so somehow the information about the original measurement is
lost when the second measurement takes place. In our neo-Everett picture we can see
fairly clearly what is going on, which we have represented here in Figure \ref{figure}.\\

A brief note is needed first on the figure itself. The worlds represented are those of the measuring device; as the
figure makes clear, more than one other world can be identified in what are given here as single lines. The
diagonal lines do \emph{not} represent the splitting of worlds (a concept alien to neo-Everett theory); they can be
thought of as, in a highly schematic way, showing the passage of time between the state in which we can identify
one measuring device-structure, and when we can see two. Note further that we have simplified
$\ket{\psi_{\theta_i}}$ as $\ket{\theta_i}$. Finally, normalisation is omitted throughout.\\

We will assume for simplicity that we are dealing with the very restricted sets of observable outcomes
$\theta_1,\theta_2,a_1,a_2$. The quantities $\alpha,\beta,\gamma,\delta,\epsilon,\zeta$ on the diagram are the
relative phases of the worlds. For example, the first decomposition comes from:
\begin{eqnarray*} \ket{R(\theta_1)} \otimes \ket{\theta_1} & = & \left( \sum_i \ket{a_i}\bra{a_i} \otimes
\id \right) \ . \bigg( \ket{R(\theta_1)} \otimes \ket{\theta_1}\bigg) \\
{} & = & \braket{a_1}{\theta_1} \ket{a_1}\ket{R(\theta_1)} + \braket{a_2}{\theta_1} \ket{a_2}\ket{R(\theta_1)}\\
{} & = & \alpha \ket{a_1}\ket{R(\theta_1)} + \beta \ket{a_2}\ket{R(\theta_1)} \end{eqnarray*}

\noindent We therefore have:\\
 $\alpha = \braket{a_1}{\theta_1}  , \beta=\braket{a_2}{\theta_1}  , \gamma=\braket{\theta_1}{a_1}  ,
\delta=\braket{\theta_2}{a_1} , \epsilon=\braket{\theta_1}{a_2} , \zeta=\braket{\theta_2}{a_2}$.\\

\noindent We prepare our state by measuring $\theta$. We then take this state as the start of our
worlds-decomposition. The measuring device is in the state $R(\theta_1)$ and the system $\ket{\theta_1}$. The state
of the system can also be identified, structurally, as a superposition of the two \ket{a} states; therefore, when a
subsequent $a$-measurement is carried out we find that we can identify two states of the measuring device --
$\ket{R(a_1)}$ and $\ket{R(a_2)}$. We now have two measurement worlds, corresponding to the $\ket{a_1}$ and
$\ket{a_2}$ worlds. Each of these worlds in turn is a superposition of the two $\ket{\theta}$ states, and so we
move to the last part of the diagram. Now if there were no decoherence at measurement, if the phase information
between worlds is not lost to the correlations with the environment, then at this stage we would simply add up all
four final states to see what our overall structure looks like. Not surprisingly, it all adds up to our initial
state $\ket{R(\theta_1)} \ket{\theta_1}$: if there is no decoherence, repeated measurement gets the initial answer
the second time around. However, with decoherence the worlds at measurement become independent of each other (for
all practical purposes) and cannot interfere. Thus we are left with four structures rather than one by the end of
the experiment, and the information about the original state we prepared is lost as this is carried in the phases
of the worlds.

\section{Conclusion}

We have analyzed the uncertainty relations in terms of a neo-Everettian picture of quantum mechanics and found that
they relate constraints on probability distributions over worlds for single systems. As a consequence we recover a
view of the impossibility of joint measurement which is invalid on traditional arguments. We have also clearly
analyzed a repeated measurement scenario in terms of neo-Everettian worlds.\\
We have found that a single world in one basis is a whole set of worlds in another, and each way of decomposing
into a worlds-system is equally valid. When the bases are related as group and generator their probability
distributions must satisfy the uncertainty relations (\ref{ur}). We have seen that a neo-Everettian interpretation
of such statistical relations is fundamentally different from that in the standard interpretation as they refer to
single systems not ensembles. This has consequences far beyond the interpretation of the uncertainty relations. To
give one example, the concept of quantum information is statistical, its definition using the density matrix of
probabilities for the state, so it will have a very different physical interpretation in a neo-Everett theory from
in one with an ensemble theory of probability.\\
 The impossibility of joint measurement and the loss of information in repeated
measurements that we have seen are not fundamentally \emph{consequences} of the uncertainty relations -- they are
consequences of the state and how the different distributions are connected. Moreover, there are other pairs of
observables not linked by uncertainty relations that are nevertheless not jointly measurable. However the relations
do link together as a single set of phenomena that which would otherwise have been disparate, by revealing
consequences of the underlying group structure. They are not fundamental, but neither are they empty of interesting
content.

\subsection*{Thanks}

Many thanks to Harvey Brown for pointing me in the direction of Uffink's work, and also for suggesting the term
`neo-Everettian' for the interpretation of quantum mechanics used here. Some of the work here was presented at the
12th Annual Foundations of Physics conference, and I am grateful for the comments and suggestions received there.

\bibliographystyle{unsrt}
\bibliography{bibliogr}

\end{document}